\documentclass[final,times,5p]{elsarticle}
\usepackage{graphicx}
\usepackage{xspace}
\usepackage{amsmath}
\usepackage{amsfonts}
\usepackage{psfrag}
\usepackage[psamsfonts]{amssymb}
\usepackage[dvips]{color}
\usepackage{srcltx}
\usepackage{hyperref}

\newcommand{\SKIP}[1]{}

\newcommand{\TCM}[1]{\textcolor{black}{#1}}

\newcommand{\PrevLet}{\textcolor{black}{the preceding Letter}}

\newcommand{\ee}{\ensuremath{e^{+}e^{-}}\xspace}

\newcommand{\PP}{\ensuremath{\psi(2S)}\xspace}
\newcommand{\JP}{\ensuremath{J/\psi}\xspace}
\newcommand{\DP}{\ensuremath{\psi(3770)}\xspace}
\newcommand{\eehadr}{\ensuremath{e^{+}e^{-}\!\!\to\text{\footnotesize hadrons}\xspace}}
\newcommand{\DD}{\ensuremath{D\overline{D}{}}\xspace}
\newcommand{\DnDn}{\ensuremath{D^{0} \overline{D}\,^{0} } \xspace}
\newcommand{\DpDm}{\ensuremath{D^{+}D^{-}}\xspace}

\newcommand{\NR}{\ensuremath{NR}\xspace}

\renewcommand{\Re}{\ensuremath{\text{Re}}\,}
\newcommand{\Rnnpm}{\ensuremath{r^{\,\text{\tiny 00}}_{\,\text{\tiny +}\text{\tiny --}}\,}}
\newcommand{\Rpmnn}{\ensuremath{r_{\,\text{\tiny 00}}^{\,\text{\tiny +}\text{\tiny --}}\,}}
\renewcommand{\Im}{\ensuremath{\text{Im}}\,}
\renewcommand{\epsilon}{\varepsilon}

\newcommand{\Gee}{\ensuremath{\Gamma_{ee}}\xspace}
\newcommand{\GeeDP}{\ensuremath{\Gamma_{ee}}\xspace}

\begin{document}
\begin{frontmatter}
  \title{
    Measurement of $\psi(3770)$ parameters
  }
\author[binp]{V.\,V.\,Anashin}
\author[binp,nsu]{V.\,M.\,Aulchenko}
\author[binp,nsu]{E.\,M.\,Baldin} 
\author[binp]{A.\,K.\,Barladyan}
\author[binp]{A.\,Yu.\,Barnyakov}
\author[binp]{M.\,Yu.\,Barnyakov}
\author[binp,nsu]{S.\,E.\,Baru}
\author[binp]{I.\,Yu.\,Basok}
\author[binp,nsu]{O.\,L.\,Beloborodova}
\author[binp]{A.\,E.\,Blinov}
\author[binp,nstu]{V.\,E.\,Blinov}
\author[binp]{A.\,V.\,Bobrov }
\author[binp]{V.\,S.\,Bobrovnikov}
\author[binp,nsu]{A.\,V.\,Bogomyagkov}
\author[binp,nsu]{A.\,E.\,Bondar}
\author[binp]{A.\,R.\,Buzykaev}
\author[binp,nsu]{S.\,I.\,Eidelman}
\author[binp]{D.\,N.\,Grigoriev}
\author[binp]{Yu.\,M.\,Glukhovchenko}
\author[binp]{V.\,V.\,Gulevich}
\author[binp]{D.\,V.\,Gusev}
\author[binp]{S.\,E.\,Karnaev}
\author[binp]{G.\,V.\,Karpov}
\author[binp]{S.\,V.\,Karpov}
\author[binp,nsu]{T.\,A.\,Kharlamova}
\author[binp]{V.\,A.\,Kiselev}
\author[binp]{V.\,V.\,Kolmogorov}
\author[binp,nsu]{S.\,A.\,Kononov}
\author[binp]{K.\,Yu.\,Kotov}
\author[binp,nsu]{E.\,A.\,Kravchenko}
\author[binp,nsu]{V.\,F.\,Kulikov}
\author[binp,nstu]{G.\,Ya.\,Kurkin}
\author[binp,nsu]{E.\,A.\,Kuper}
\author[binp,nstu]{E.\,B.\,Levichev}
\author[binp]{D.\,A.\,Maksimov}
\author[binp]{V.\,M.\,Malyshev}
\author[binp]{A.\,L.\,Maslennikov}
\author[binp,nsu]{A.\,S.\,Medvedko}
\author[binp,nsu]{O.\,I.\,Meshkov}
\author[binp]{S.\,I.\,Mishnev}
\author[binp]{I.\,I.\,Morozov}
\author[binp]{N.\,Yu.\,Muchnoi}
\author[binp]{V.\,V.\,Neufeld}
\author[binp]{S.\,A.\,Nikitin}
\author[binp,nsu]{I.\,B.\,Nikolaev}
\author[binp]{I.\,N.\,Okunev}
\author[binp,nstu]{A.\,P.\,Onuchin}
\author[binp]{S.\,B.\,Oreshkin}
\author[binp]{I.\,O.\,Orlov}
\author[binp]{A.\,A.\,Osipov}
\author[binp]{S.\,V.\,Peleganchuk}
\author[binp,nstu]{S.\,G.\,Pivovarov}
\author[binp]{P.\,A.\,Piminov}
\author[binp]{V.\,V.\,Petrov}
\author[binp]{A.\,O.\,Poluektov}
\author[binp]{V.\,G.\,Prisekin}
\author[binp]{A.\,A.\,Ruban}
\author[binp]{V.\,K.\,Sandyrev}
\author[binp]{G.\,A.\,Savinov}
\author[binp]{A.\,G.\,Shamov\corref{cor}}
\author[binp]{D.\,N.\,Shatilov}
\author[binp,nsu]{B.\,A.\,Shwartz}
\author[binp]{E.\,A.\,Simonov}
\author[binp]{S.\,V.\,Sinyatkin}
\author[binp]{A.\,N.\,Skrinsky}
\author[binp]{V.\,V.\,Smaluk}
\author[binp]{A.\,V.\,Sokolov}
\author[binp]{A.\,M.\,Sukharev}
\author[binp,nsu]{E.\,V.\,Starostina}
\author[binp,nsu]{A.\,A.\,Talyshev}
\author[binp,nsu]{V.\,A.\,Tayursky}
\author[binp,nsu]{V.\,I.\,Telnov}
\author[binp,nsu]{Yu.\,A.\,Tikhonov}
\author[binp,nsu]{K.\,Yu.\,~Todyshev\corref{cor}}
\cortext[cor]{Corresponding authors, e-mails: \\ shamov@inp.nsk.su,~~todyshev@inp.nsk.su }
\author[binp]{G.\,M.\,Tumaikin}
\author[binp]{Yu.\,V.\,Usov}
\author[binp]{A.\,I.\,Vorobiov}
\author[binp]{A.\,N.\,Yushkov}
\author[binp]{V.\,N.\,Zhilich}
\author[binp,nsu]{V.\,V.\,Zhulanov}
\author[binp,nsu]{A.\,N.\,Zhuravlev} 

\address[binp]{Budker Institute of Nuclear Physics, Siberian Div., 
  Russian Acad. Sci., 630090, Novosibirsk, Russia}
\address[nsu]{Novosibirsk State University, 630090, Novosibirsk, Russia}
\address[nstu]{Novosibirsk State Technical University, 
  630092, Novosibirsk, Russia}

\begin{abstract}
We report the final results of a study of the $\psi(3770)$ meson 
using a data sample collected  with the KEDR detector at the VEPP-4M 
electron-positron collider. The data
analysis takes into account interference
between the resonant and nonresonant \DD production, where
the latter is related to the nonresonant part of the 
energy-dependent  form factor $F_D$.
The vector dominance approach and several 
empirical parameterizations have been tried for the nonresonant $F_D^{\NR}(s)$.

Our results for the mass and total width of \DP are
\begin{equation*}
 \begin{split}
  M & =  3779.2\,\,^{+1.8}_{-1.7}\,\, ^{+0.5}_{-0.7} \,\,^{+0.3}_{-0.3} \,\, \text{MeV}, \\
 \Gamma & = \:\:\:\:\,24.9\,\,^{+4.6}_{-4.0}\,\, ^{+0.5}_{-0.6}\,\,^{+0.2}_{-0.9} \,\,    \text{MeV,} \\         
 \end{split}
\end{equation*}
where the first, second and third uncertainties 
are statistical, systematic and model, respectively. 
For the electron partial width two possible solutions have been found:
\begin{equation*}
  \begin{split}
& (1)  \:\:\:\:\: \GeeDP = 154\,\,^{+79}_{-58}\,^{+17}_{-9}\,^{+13}_{-25}\: \text{eV},\\
& (2)  \:\:\:\:\: \GeeDP = 414\,\,^{+72}_{-80}\,^{+24}_{-26}\,^{+90}_{-10}\: \text{eV}.
  \end{split}
\end{equation*}
Our statistics are insufficient to prefer one solution to another.
The solution (2) mitigates the problem of non-\DD
decays but is disfavored by potential models.

It is shown that  taking into account
the resonance--continuum interference
in the near-threshold region affects resonance parameters, 
thus the results presented 
can not be directly compared with the corresponding PDG values obtained 
ignoring this effect.
\end{abstract}
\end{frontmatter}
\section{Introduction}
\label{sec:intro}

The preceding Letter of this volume
is devoted to the measurement
of the \PP meson parameters in the KEDR experiment performed during
energy scans from 3.67 to 3.92 GeV at the VEPP-4M \ee collider.
In this Letter we describe the application of the developed
tools to the measurement of \DP parameters
omitting details common for \PP and \DP.

Since the discovery of the \DP, seven experiments contributed to 
the determination of its parameters, nevertheless the situation 
with the mass, total width and electron
partial width is still not clear.

The incomplete compilation of results
reported on \DP mass is presented in Table~\ref{Tab:mComp}.
It does not include the results of Refs.~\cite{Li:2009pw,Zhang:2009gy}
with the analysis of the $\ee\!\to\DD\,$ data of BES~\cite{Ablikim:2008zzDD} 
and the $\ee\!\to\DD\gamma\,$ data of Belle~\cite{Pakhlova:2010ek}
in which  the \DP electron width has been fixed in the fits causing a
mass bias. In addition,  the bin size in Belle data around \DP 
seems too large for a simple center-of-bin fitting.
These works encouraged us to employ the vector dominance model in 
the analysis~\cite{Todyshev:2010zz}.

The values presented form
three partially overlapping clusters. The first one with
$\left<M\right>\!=\!3772.5\pm0.4~{\rm MeV}$ comes from the analyses 
in which interference between resonant and nonresonant \DD production 
has been
ignored~\cite{DPDiscSLAC,DLCPSIPP,MRK2PSIPP,Todyshev:2005zp,Ablikim:2006md,Ablikim:2007gd}.
\TCM{In addition, the analyses assumed the simplest shape of nonresonant
\DD--cross section similar to that for point-like pseudoscalars in QED.}
The statistical uncertainty in this case is small
(in \cite{Ablikim:2007gd} the influence of $\psi(4040)$ and higher $\psi$'s
included in the analysis increases the \DP mass uncertainty).
The second cluster of $B\to D\overline{D}{}K$ analyses~
\cite{Abe:2003zv,:2007aa,Aubert:2007rva} has 
$\left<M\right>\!=\!3775.6\pm2.3~{\rm MeV}$
(the result of \cite{Abe:2003zv}  is not included because 
of its uncertain status).
The third, highest mass, cluster is formed by the analyses 
accounting for interference~\cite{BABARDATA,Todyshev:2010zz}
 and gives $\left<M\right>\!=\!3777.3\pm1.3~{\rm MeV}$.

As was mentioned in Section 5.2 of the previous Letter, taking into 
account the resonance--continuum interference is essential for a 
determination of the \DP parameters. A close \DD production threshold
significantly increases the importance of that.
A consideration of the interference effects
is one of the primary goals of this 
experiment\footnote{The result of~\cite{Todyshev:2005zp} was
obtained solely to check consistency with the previous measurements.}.
\SKIP{It should be noted that consideration of the effects of 
resonance -- continuum interference near the \,$D\overline{D}\,$
production threshold is essential for the determination of the
\DP\, parameters and is one of the primary goals of this 
experiment\footnote{The result of~\cite{Todyshev:2005zp} was
obtained solely to check the agreement with the previous measurements.}.}

If interference is ignored  in a fit of the measured
\DD\, or multihadron  cross section,
a bias appears in the growing continuum contribution that causes
a bias in the resonance amplitude and a shift of the mass value.
The signs of these effects depend on the relative position of the interference
peak and dip. The \DD cross section at the
threshold is fixed at zero, 
therefore the weights of the more distant data points
in a fit are larger than those of the less distant ones.
Evidence for a dip after the \DD cross section maximum is visible 
in all published data with large enough statistics (see, for example, 
Fig.~1 of Ref.~\cite{Ablikim:2006md}), therefore, 
the artificial mass shift should be negative (undercounted
events move the resonance peak to the left). That is exactly 
what we observe analyzing the published mass results.

If the result on mass of~\cite{Abe:2003zv} is ignored, 
the \DP mass value obtained in $B$ decays does not contradict neither 
to $3772.5$ nor $3777.3~{\rm MeV}$.
The interference of the resonant and nonresonant $D\overline{D}$ yields
also takes place in this case but the relation between them can differ 
from that in \ee collisions, besides, the interference effect can be 
partially compensated by subtraction of the combinatorial background. 
Thus, the intermediate mass value does not seem surprising.

Below we briefly describe the theoretical basis of the analysis performed,
enter some details concerning the analysis procedure 
and not covered in \PrevLet, present the results on the
\DP parameters and discuss their systematic uncertainties
and model dependence.
\renewcommand{\arraystretch}{1.1}
\begin{table}[t]
\caption{\label{Tab:mComp} Incomplete compilation of results on $\psi(3770)$
mass.}
\begin{center}
\begin{tabular}[c]{|l|l|l|} \hline
Analysis &   $M_{\DP}$[MeV]      & Comments \\ \hline
MARK-I \cite{DPDiscSLAC}  & $3774.1 \:\:\:\:\:\:\pm 3$  & $\eehadr\,^{(a)}$ \\  \hline
DELCO \cite{DLCPSIPP}   & $3772.1 \:\:\:\:\:\:\pm 2$  & $\eehadr\,^{(a)}$  \\  \hline
MARK-II \cite{MRK2PSIPP} & $3766.1 \:\:\:\:\:\:\pm 2$ & $\eehadr\,^{(a)}$   \\  \hline
Belle \cite{Abe:2003zv} & $3778.4 \pm 3.0 \pm 1.3 $ 
                 &  $B\to D^0\overline{D}\,^0K^{+}\,^{(b)}$\\  \hline
KEDR \cite{Todyshev:2005zp}& $3773.5\pm 0.9 \pm 0.6 $ & $\eehadr\,^{(c)}$\\  \hline 
BES-II \cite{Ablikim:2006md} &$3772.4 \pm 0.4 \pm 0.3$  & $\eehadr\,^{(a)}$ \\   \hline
BES-II \cite{Ablikim:2007gd} & $3772.0\:\:\:\:\:\:\pm 1.9 $      & \eehadr \\   \hline
Belle\cite{:2007aa} & $3776.0 \pm 5.0 \pm 4.0 $&  $B\to D^0\overline{D}\,^0 K^{+}$\\   \hline
BaBar \cite{Aubert:2007rva} & $3775.5 \pm 2.4 \pm 0.5 $    &   $B\to D\overline{D}K$\\   \hline
BaBar \cite{BABARDATA}  & $3778.8 \pm 1.9 \pm 0.9 $&
                       $e^{+}e^{-}\!\!\to D\overline{D}\gamma\,^{(d)}$ \\ \hline
\SKIP{
  \cite{Li:2009pw}, data of &$3776.0 \pm 1.0 \:\: \pm \:\! \text{?}$
                       & $e^{+}e^{-}\!\!\to D\overline{D}\:+$ \\
     BES+Belle &  &      $e^{+}e^{-}\!\!\to D\overline{D}\gamma\,^{(d,e,g)}$ \\\hline
   \cite{Zhang:2009gy}, BES data& $3774.0 \pm  1.0 \:\: \pm \:\! \text{?}$
                            & $e^{+}e^{-}\!\!\to D\overline{D}\,^{(d,e,f,g)}$ \\ \hline
}
KEDR \cite{Todyshev:2010zz} & $3778.0 \pm 1.6 \pm 0.7$ & $\eehadr\,^{(c,d)}$ \\ \hline
\multicolumn{3}{l}{}\\
\multicolumn{3}{l}{$\,^{(a)\:}${\footnotesize --- omitted in the latest 
PDG edition}}\\
\multicolumn{3}{l}{$\,^{(b)\:}${\footnotesize --- the result on 
${\cal B}(B\to D^0\overline{D}\,^0 K^{+})$ is superseded by \cite{:2007aa}}}\\
\multicolumn{3}{l}{$\,^{(c)\:}${\footnotesize --- preliminary results 
reported at various conferences}}\\
\multicolumn{3}{l}{$\,^{(d)\:}${\footnotesize --- interference between resonant and nonresonant}} \\
\multicolumn{3}{l}{\phantom{$\,^{(e)\:}${\footnotesize --- }}{\footnotesize $D\overline{D}$ production is taken into account}}\\
\SKIP{
\multicolumn{3}{l}{$\,^{(e)\:}${\footnotesize --- the authors analyze the published data,}}\\
\multicolumn{3}{l}{\phantom{$\,^{(d)\:}${\footnotesize ---}}{\footnotesize systematic
\TCM{uncertainties} are not evaluated}}\\
\multicolumn{3}{l}{$\,^{(f)\:}${\footnotesize --- analysis of BELLE data is not included, the bin size seems}}\\
\multicolumn{3}{l}{\phantom{$\,^{(f)\:}${\footnotesize --- }}{\footnotesize too wide to be used alone for the \DP mass determination}}\\
\multicolumn{3}{l}{$\,^{(g)\:}${\footnotesize --- electron width of \DP is fixed at the world average value}}\\
\multicolumn{3}{l}{\phantom{$\,^{(g)\:}${\footnotesize --- }}{\footnotesize causing a negative bias in the mass}}\\
}
\end{tabular}
\end{center}
\end{table}

\section{Multihadron cross section in the vicinity of \DP}

A few approaches can be employed to determine the resonance
parameters using a multihadron cross section data. In the
\TCM{Ref.}~\cite{Ablikim:2006md} the fit of the $R$ ratio was performed,
in the \TCM{Ref.}~\cite{Ablikim:2008zzcAnShp} the efficiency--corrected cross section
was analyzed. There are many different sources of multihadron events 
such as the \PP and \DP production, the light quark production etc., thus
the variation of the net detection efficiency
in the whole experiment range can exceed 20\%~\cite{DZhang}. The calculation
of the net efficiency implies knowledge of the resonance parameters
and accounting for the interference effects, therefore
an iterative analysis is required. In this work we fit the observed
multihadron cross section not corrected for the detection efficiency which
allows iterations to be avoided.

\subsection{Observed cross section and $D$-meson form factor}

In the energy range from slightly below the $\psi(2S)$ peak to slightly above 
the $\DD\pi$ threshold the variation of the light quark contribution 
to $R$ ($R_{uds}$) is small, so that the multihadron
cross section observed in the experiment can be written as
\begin{equation}
\begin{split}
& \sigma^{\mathrm{obs}}_{mh} = 
     \epsilon_{\psi(2S)}\, \sigma^{RC}_{\psi(2S)} +
     \epsilon_{J/\psi}\, \sigma^{RC}_{J/\psi} +
     \epsilon_{\tau\tau}\,\sigma^{RC}_{\tau\tau} +
     \sigma^{emp}_{uds} \,+\\
    & \epsilon_{\DpDm}\,\sigma^{RC}_{\DpDm} + 
      \epsilon_{\DnDn}\,\sigma^{RC}_{\DnDn} +
      \epsilon_{n\DD}\,\mathcal{B}_{n\DD}\,
           \sigma^{RC}_{\DP} \,+ \\ &\sigma^{emp}_{\DD\pi}
\end{split}
\label{eq:SigMHobs}
\end{equation}
where $\sigma^{RC}$'s are theoretical
cross sections, $\epsilon$'s are corresponding detection efficiencies,
and $\sigma^{emp}$'s are terms treated empirically
as described below. The $RC$ superscript means that the cross
section has been corrected 
for initial state radiation (ISR) effects,
$n\DD$ stands for the direct \DP decay to light
hadrons, the other (super/sub)scripts seem self-explanatory,
$\mathcal{B}_{n\DD}$ is a branching fraction.
All detection efficiencies explicitly entering 
Eq.~\eqref{eq:SigMHobs} can be kept energy independent
with sufficient accuracy for the event selection criteria employed
(see Sec.~\ref{sec:DetEff}).

The first four terms have no
peculiarities in the whole energy range of the experiment, while
the last four are responsible for
the excess of the cross section in the \DP region.

The fourth term of Eq.~\eqref{eq:SigMHobs}
corresponding to the light
quark contribution can be scaled as $1/s^{1-\delta}$
where a relatively small parameter $\delta$ is due to 
the energy dependence of the detection efficiency
and radiative corrections. Possible variation of $R_{uds}$
can also contribute to $\delta$.
This term can be easily removed from the consideration in the fit
of the cross section provided that the $\delta$
value is known. 
The $\DD\pi$ cross section can be treated as a small correction.
We took it into
account using the approximately known shape and an additional fit parameter.

Calculations for 
$\sigma^{RC}_{\psi(2S)}$ and $\sigma^{RC}_{\tau\tau}$ are
described in \PrevLet,
a small contribution of the $J/\psi$ tail was calculated
similarly to the \PP one,
for the \DD production cross section (here and below $D$ stands
for $D^{+}$ or $D^0$) one has
\begin{equation}
\begin{split}
\sigma^{RC}_{\DD}(W) = & \!\int\!
z_{\DD}\left(W^{\prime}\sqrt{1\!-\!x}\,\right)\,
 \sigma_{\DD}\left(W^{\prime}\sqrt{1\!-\!x}\,\right) \\
   & \times\mathcal{F}(x,W^{\prime\,2})\,G(W,W^{\prime})\,\,dW^{\prime}dx,
\end{split}
\end{equation}
where $\mathcal{F}(x,s)$ is the probability to lose a fraction of $s$
in the initial state radiation~\cite{KF}, $G(W,W^{\prime})$ describes
a distribution of the total collision energy, 
which can be assumed to be Gaussian with an
energy spread $\sigma_{W}$.

For the charged mode ($D^+D^-$) the factor $z_{\DpDm}$ describing the
Coulomb interaction between the mesons produced~\cite{Voloshin:Charmonium}
is taken according to 
Sommerfeld-Sakhar\-ov~\cite{SommerfeldRef,SakharovRef,MiltonSolovtsovFormula}:
\begin{equation}
z_{\DpDm} = \frac{\pi\alpha/\beta_{D^{+}}}{1-\exp{(-\pi\alpha/\beta_{D^{+}})}}
\times \theta(W\!-\!2 m_{D^{+}}).
\label{eq:Zpm}
\end{equation}
For the neutral mode ($\DnDn$) there is no such interaction, thus
\begin{equation}
z_{\DnDn} = 1\times \theta(W\!-\!2 m_{D^{0}}),
\label{eq:Znn}
\end{equation}
the step functions $\theta(W\!-\!2 m_{D})$ 
are shown explicitly to simplify some expressions below.

The cross section $\sigma_{\DD}$
can be expressed via the form factor $F_D$ and $D$-meson velocity 
in the c.m.system $\beta_D$:
\begin{equation}
\sigma_{\DD}(W) =\frac{\pi\alpha^2}{3W^2}\,\, \beta_D^{3}\, \left|F_D(W)\right|^2,
 \:\:\:\: \beta_D = \sqrt{1-4m_D^2/W^2}.
\label{eq:xsDD}
\end{equation}

To determine the parameters of resonances above the \DD threshold, their
amplitudes should be separated in $F_D$:
\begin{equation}
\label{eq:FF}
F_D(W) =  \sum\limits_i F^{R_{i}}_D(W)\,e^{i\phi_i} + F^{\NR}_D(W),
\end{equation}
where $\phi_i$ is the phase of the $i$-th resonance $R_i$ 
relative to $F^{\NR}_D$.
 
For the resonance with the partial widths $\Gamma_{ee}$ and $\Gamma_{\DD}$ and
the total width $\Gamma(W)$, one has a Breit-Wigner amplitude
\begin{equation}
F^{R}_D(W) = \frac{6\sqrt{\left(\Gamma_{ee}/\alpha^2\right)\,
    \left(\Gamma_{\DD}(W)/\beta_D^3\right)\,}\,W}
                            {M^2-W^2-iM\Gamma(W)}
\label{eq:BW}
\end{equation}
(the vacuum polarization factor is included in $\Gamma_{ee}$).

Considering $\Gamma(M)$ as a nominal resonance width and introducing
the sum of the branching fractions to all non-\DD modes
$\mathcal{B}_{n\DD}$, one obtains the energy-dependent \DD partial width
\begin{equation}
\Gamma_{\DD}(W) =
\frac{ (M/W)\,\,z_{\DD}(W)\,d_{\DD}(W)
\cdot\Gamma(M)\cdot(1-\mathcal{B}_{n\DD})}{z_{\DnDn}(M)d_{\DnDn}(M)+
  z_{\DpDm}(M)\,d_{\DpDm}(M)}
\label{eq:GammavVSw}
\end{equation}
in line with the PDG prescriptions (p. 808 of Ref.~\cite{PDG}).
Here $d_{\DpDm}$ and $d_{\DnDn}$
are the Blatt-Weisskopf damping factors 
for a  vector resonance~\cite{BlattWeisskopf}:
\begin{equation}
 d_{\DD} = \frac{\rho_{\DD}^3}{\rho_{\DD}^2+1}, \:\:\:\:
\rho_D = q_D R_0,
\label{eq:BWdampF}
\end{equation}
where $R_0$ represents the meson radius and $q_D$ is the c.m. momentum of the
meson $q_D =\beta_D\,W/2$. The partial width dependence
according to Eq.~\eqref{eq:GammavVSw} corresponds to the approach of
Ref.~\cite{Voloshin:Charmonium}. Its simplified form was used
in the experiments~\cite{DPDiscSLAC,DLCPSIPP,MRK2PSIPP}. The approach
is somewhat different from that employed in
\TCM{Refs.~\cite{Ablikim:2006md,Ablikim:2007gd} by BES}
which does not lead
to noticeable changes of the \DP parameters.

The $\DD\pi$ cross section entering (\ref{eq:SigMHobs}) as a small
correction can be calculated with sufficient accuracy using
\begin{equation} 
\begin{split}
&
\sigma_{\DD\pi}(W) =\frac{\pi\alpha^2}{3W^2}\,\, \beta_{\DD\pi}^{3}\,
        \left|F_{\DD\pi}\right|^2, \\[1pc]
&\beta_{\DD\pi} =\sqrt{(1\!-\!(m_{D^{*}}\!+\!m_D)^2/W^2)
 (1\!-\!(m_{D^{*}}\!-\!m_D)^2/W^2)}\,.
\end{split} 
\label{eq:CMSbeta}
\end{equation} 
The quantity $F_{\DD\pi}$ is treated as a fit parameter.
\subsection{Nonresonant $D$-meson form factor}\label{sec:Fnr}
The nonresonant part of the form factor can be written as
\begin{equation}
F^{\NR}_D(W) = \frac{1}{|1-\Pi_0(W)\,|} \,f_D(W)
\label{eq:FdNR}
\end{equation}
with $f_D(W)=1$ for point-like particles. Here $\Pi_0$ is the 
vacuum polarization operator
except the contributions of all resonances which are written
separately in (\ref{eq:FF}). 
We remind that the full polarization operator
is calculated using the total cross section of $e^+e^- \to hadrons$
that already includes all resonances, therefore use 
of the full operator $\Pi$ instead of $\Pi_0$ in the nonresonant amplitude
leads to double counting of the
resonances and thus incorrect values of the leptonic widths (see also the
discussion in Sec. 5.3 of \PrevLet).

There are no precise theoretical predictions for $F^{\NR}_D(W)$. 
The model-independent
result can be obtained using the expansions of $\Re F^{\NR}_D(W)$ and
$\Im F^{\NR}_D(W)$ at the point $W\!=\!M$ with the coefficients 
free in the
fit. Our statistics are not sufficient for that, thus we
have to rely on some model or use a pure empirical
approach as in  \TCM{Ref.~\cite{BABARDATA} by BaBar}
also taking into account the resonance--continuum interference.

The most certain prediction of the form factor can
be obtained with an application of the \, Vector Dominance \, Model\, (VDM)
to charm production. Standard VDM assumes that the inclusive
cross section $e^{+}e^{-} \to hadrons$ at low energy is saturated
by the interfering contributions of the limited number of vector mesons.
A similar assumption can be accepted for the inclusive
$e^{+}e^{-}\to c\overline{c}$ cross section and its
exclusive modes such as $e^{+}e^{-}\to D\overline{D}$. The VDM--like
analysis  of the $R$ ratio in the energy
range of $W=3.7\div5$~GeV has been performed by BES in
Ref.~\cite{Ablikim:2007gd}, where the light quark contribution  $R_{uds}$
was calculated using pQCD. The work cited accounts for
$\psi(3770)$, $\psi(4040)$, $\psi(4160)$ and $\psi(4415)$ resonances
but does not account for a possible contribution of \PP decays to
\DD above the threshold. 
\SKIP{This contribution has been studied in
Refs.~\cite{Li:2009pw,Zhang:2009gy} already mentioned in the introduction
to this Letter. The studies include a theoretical consideration and
some analysis of the \DD cross section measured by BES as well BELLE.
}Studies of this contribution Refs.~\cite{Li:2009pw,Zhang:2009gy} 
include a theoretical consideration and
some analysis of the \DD cross section measured by BES as well as by BELLE.
In this work we employ VDM in a simplified form
\begin{equation}
  F^{\NR}_D(W) = F^{\PP}_D(W) + F_0,
\label{eq:PPvdm}
\end{equation}
where $F_0$ is a real constant representing the contributions of the
$\psi(4040)$ and higher $\psi$'s.
The \PP contribution to the \DD form factor
$F^{\PP}_D$ was calculated using Eq.~\eqref{eq:BW} with the
\DnDn and \DpDm partial widths defined similarly
to Eq.~\eqref{eq:GammavVSw} with a specific value of the
effective radius $R_0$.  The value of
$\Gamma^{\psi(2S)}_{D\overline{D}}(M_{ref})\!=\!
\Gamma^{\psi(2S)}_{\DpDm}(M_{ref}) + \Gamma^{\psi(2S)}_{\DnDn}(M_{ref})$
at some reference point $M_{ref}$, as well as the constant $F_0$,
should be obtained from the data fit ($M_{ref}\!=\!3778$~MeV
was used). The partial width ratio
$\Gamma^{\psi(2S)}_{\DnDn}/\Gamma^{\psi(2S)}_{\DpDm}$
is presumably close to that of $\psi(3770)$.

To evaluate the model dependence of the \DP parameters
we tried a few nonresonant form factor parameterizations, which do not
assume vector dominance. The most popular empirical parameterization is
 probably exponential:
$$f_D=\exp{(-q_D^2/a^2)}\,,$$
where $q_D$ is the c.m. momentum~\cite{FormFactor1976}.
It is well motivated 
far above the threshold but has few parameters to describe the low 
energy region.
Instead of it we used
\begin{equation}
  f_D = -\frac{g_q}{(1+a_q\,q_D^2+b_q\,q_D^4)^n}\:\:\:\:\:\:(n=0.5,1).
  \label{eq:FnrQ}
\end{equation}
The minus sign is chosen to match the \PP dominance expectations.
In the case $n\!=\!0.5$, $b_q=0$, the nonresonant cross section acquires the
Blatt-Weisskopf factor \eqref{eq:BWdampF} with $R_0=a_q$. The case
$n\!=\!1$ corresponds to a more rapid form factor fall. Use of two
parameters $a_q$ and $b_q$ allows us to take into account 
in the limited energy range 
the increase of the \DD cross section described by the $G(3900)$ structure in
the \TCM{Ref.~\cite{BABARDATA}.}
Alternatively, the dependence on $W\!-\!m_D$
\begin{equation}
  f_D = -\frac{g_W}{1+a_W (W\!-\!2 m_D)+b_W (W\!-\!2 m_D)^2}
  \label{eq:FnrW}
\end{equation}
and combined dependences
\begin{equation}
  f_D  = -\frac{g_{qW}}{(1+ a_{qW}\, (W\!-\!2m_D) + b_{qW}\,q_D^2)^n}
  \label{eq:FnrQW}
\end{equation}
were considered.

To check validity
of the \PP domination hypothesis in Eq.~\eqref{eq:PPvdm}
the following parameterizations were used:
\begin{equation}
  f_D  = \frac{g_m}{a_m-W}\, \left(1+\frac{i\,b_m\,\beta_D^n}{a_m-W}\,\right) 
\:\:\:\:\:\:(n=0,1,3)\,,
  \label{eq:FnrBetaExp}
\end{equation}
where $\beta_D$ is the $D$--meson velocity.
They are expansions of the Breit-Wigner amplitude
with the mass $a_m$ treated as a free parameter, the values of $n$
correspond to different assumptions on $\Gamma(W)$ dependence.
In case of \PP dominance the fitted value of $a_m$ would be close to
$M_{\PP}$.

\section{Data analysis}
\subsection{Detection efficiency determination}
\label{sec:DetEff}

To perform a fit of the observed multihadron
cross section with Eq.~\eqref{eq:SigMHobs}, it is necessary to
know six detection efficiencies explicitly entering the equation
and the detection efficiency $\epsilon_{uds}$ implicitly contained
in the term $\sigma^{emp}_{uds}$ related to the 
continuum light quark production.
 They were determined from Monte Carlo
simulation.
The efficiency $\epsilon_{n\DD}\,$ enters  Eq.~\eqref{eq:SigMHobs} in
the product with the non-\DD branching fraction $\mathcal{B}_{n\DD}$,
which is rather uncertain. That allows one to assume
$\epsilon_{n\DD} \approx \epsilon_{\PP}$.

The event selection criteria, which are different
for 2004 and 2006 scans, and the procedure of the detection
efficiency determination for the
$\psi(2S)$ decay simulation are described in detail in \PrevLet.
The tuned version of the BES generator~\cite{BESGEN} was employed
to the obtained \PP detection efficiency in the vicinity of the peak.
The same version of the generator with parameters optimal for \PP
simulation was used to simulate the \PP and \JP tails and the
continuum $uds$ production.
To simulate $\ee\!\!\to\DD$ events, \DD pairs were first generated 
with the proper angular distribution. Decays of $D$ mesons were simulated
using the routine {\it LU2ENT} of the  \it JETSET \rm 7.4 
package~\cite{JETSET}.
The decay tables of \it JETSET \rm were updated according to 
those of the PDG review~\cite{PDG}.

\begin{table}[h!]
\caption{\label{tab:def}{\normalsize Detection efficiency for the processes
 of interest and its variation in the experiment energy range 
$\Delta W \approx 200$~MeV.}}
\begin{center}
\begin{tabular}{l|c|c|c} \hline
Process & $\epsilon_{2004}$ & $\epsilon_{2006}$ & 
     $\Delta\epsilon/\epsilon$, \%  \\ \hline
\DpDm & $0.75\pm0.02$ & $0.84\pm0.02$ & $+1.0\pm0.3$ \\ \hline
\DnDn & $0.74\pm0.02$ & $0.81\pm0.02$ & $+1.0\pm0.3$ \\ \hline
\PP & $0.63\pm0.01$ & $0.72\pm0.01$   & $-0.1\pm0.1$ \\ \hline
\JP & $0.50\pm0.02$ & $0.60\pm0.02$   & $-0.2\pm0.1$ \\ \hline
$uds$ & $0.55\pm0.02$ & $0.69\pm0.02$   & $+2.1\pm0.5$ \\ \hline
\end{tabular}
\end{center}
\end{table}

The detection efficiencies for the processes
of interest and their energy variations are presented in
Table~\ref{tab:def}. The systematic uncertainties of the efficiencies
$\epsilon_{\JP}$ and $\epsilon_{uds}$
were estimated by variation of JETSET parameters
preserving the mean value of the charged multiplicity.
The systematic \TCM{uncertainties} on $\epsilon_{\DpDm}$ and $\epsilon_{\DnDn}$
were found modifying the decay branching fractions of $D$--mesons
within uncertainties quoted in the PDG tables.

\subsection{Fitting of data}
\label{sec:fitting}

The observed multihadron cross section
was fitted as a function of $W$  with the
expression (\ref{eq:SigMHobs}) using some assumptions about
the behaviour of the nonresonant form factor $F^{\NR}_D$.
The details on the likelihood calculation can be found in 
\PrevLet.
The following additional constraint was applied
\begin{equation}
 \left|\frac{F^{\NR}_{D^{+}}(W_{ref})}{F^{\NR}_{D^{0}}(W_{ref})}\right|^2 = \:
 \frac{\sigma_{\DpDm}(W_{ref})}{\sigma_{\DnDn}(W_{ref})} = \:
 \Rpmnn\,,
\end{equation}
with the reference mass $W_{ref}\!=\!3773$~MeV not far from
the observed cross section maximum.
The value
$\Rnnpm = 0.776^{+0.028}_{-0.025}$~\cite{CLEOrPM00}
was used. The world average values  were also
used for the \JP mass, total and electronic width. The total width
of \PP was fixed at the value of $296\pm9$~keV obtained
in \PrevLet.
The meson radii of Eq.~\eqref{eq:BWdampF} were fixed at 1~fm and
0.75~fm for \DP and $\PP$,
respectively (Refs.~\cite{Eichten:1980,Buchmuller:1981,Godfrey:1985}).
Since the experimental results on the non-\DD fraction of \DP decays
$\mathcal{B}_{n\DD}$ are controversial
and theory expects  it to be small, we performed the fits with
$\mathcal{B}_{n\DD} = 0$ and $0.16$ and assigned variation of the parameters
to the systematic uncertainties.

The light quark contribution
was parameterized as
\begin{equation}
  \sigma^{emp}_{uds} = \epsilon_{uds} \,
\left(1+\delta^{RC}_{uds}\,\right)\,\overline{R}_{uds}\,
       \left(\frac{M_{\PP}^2}{s}\right)^{1-\delta}\!\!\!\!\!
    \sigma^{B}_{\mu\mu}(M_{\PP})\,,
\label{eq:1s}
\end{equation}
where $\delta^{RC}_{uds}$ is a radiative correction of  about 0.12,
$\,\overline{R}_{uds}$ is a light quark contribution to the $R$ ratio averaged
over the experiment energy range and
$\sigma^{B}_{\mu\mu}$ in a Born level dimuon cross section.
The values of $\delta^{RC}_{uds}$ and $\epsilon_{uds}$
are constants corresponding to $W\!=\!M_{\PP}$.
The parameter $\delta$ was fixed at $0.187\pm 0.046$ with the \TCM{uncertainties}
dominated by \TCM{that} of the detection efficiency variation
presented in Table~\ref{tab:def}. The detailed discussion can be found below
in Sec.~\ref{sec:systerr}.

A simultaneous fit of three scans has been performed. Each scan
has its own free parameters (the energy spread $\sigma_W$ 
and $\overline{R}_{uds}$)
 and has other free parameters common for all three scans.
Among them are
the mass $M_{\PP}$, the product of the electron width and the
branching fraction of its decay to hadrons
$\Gamma_{ee}\!\times\!{\cal B}_{hadr}$ for \PP;
the mass $M$, the total width $\Gamma$, the electron width
$\Gamma_{ee}$ and the interference phase $\phi$ for \DP. 
The $\DD\pi$ contribution was tuned using the free parameter $F_{\DD\pi}$.
The nonresonant form factor has been controlled by either the free parameters
$\Gamma^{\PP}_{\DD}(M_{ref})$ (the \PP partial width above
the \DD threshold) and $F_0$ (constant term of the form factor)
or by three parameters $g$, $a$, $b$ defined in
Eqs. \eqref{eq:FnrQ}, \eqref{eq:FnrW}, \eqref{eq:FnrQW} and
\eqref{eq:FnrBetaExp}. The last but not least free parameter
was the interference phase $\phi$. The total number of free parameters
was either 15 or 16.

The parameters controlling the nonresonant form factor
behaviour have strongly correlated asymmetric statistical errors.
Instead of them we present below the value of the nonresonant \DD 
cross section
at the resonance peak $\sigma^{NR}_{\DD}(M)$ and its error
obtained in fits with modified sets of free parameters
(e.g., the $(F_0$, $\Gamma^{\PP}_{\DD})$ pair was replaced with the
$(F_0$, $\sigma^{NR}_{\DD})$ one).
\begin{figure}[t!]
\includegraphics[width=0.48\textwidth]{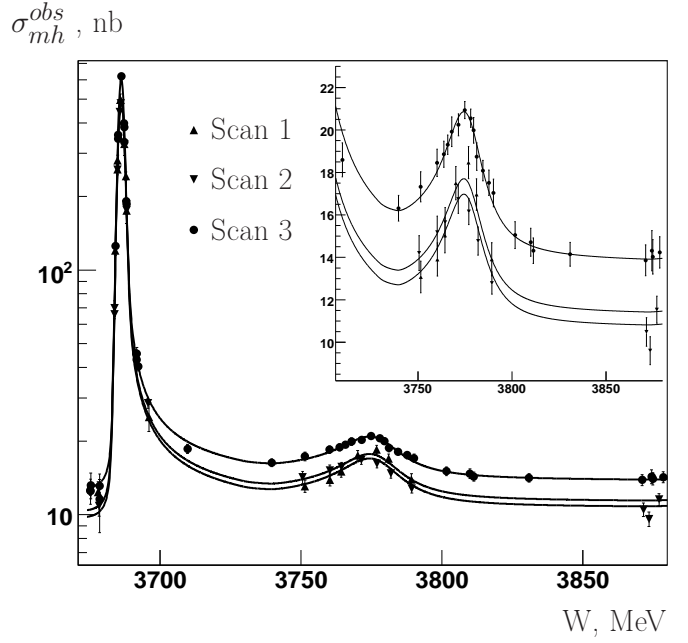}
\caption {\label{ScansFig} \normalsize{
 The observed multihadron cross section as a function of the c.m. energy
 for the three scans. The curves are the results of the vector dominance fit.
 The detection efficiencies and the energy spreads for the scans differ.
}}
\end{figure}

The observed multihadron cross section for the scans
is presented in Fig.~\ref{ScansFig}. The curve represents the
vector dominance fit.
The resulting values of \PP parameters agree very well with those
obtained fitting the narrow energy range
around \PP (previous Letter). The difference in the mass
values is 2 keV, the variation of the $\Gamma_{ee}\times\mathcal{B}_h$
product is about 0.3\%.
As a consistency check, we estimate
$\overline{R}_{uds}$ for the three scans.
The fitted values
are $2.33\pm 0.10$, $2.25\pm 0.09$
and $2.31\pm0.06$. The weighted average
$\overline{R}_{uds} = 2.300 \pm 0.046 \pm 0.108$ ($\chi^2/N_{DoF}=0.49/2$)
agrees well with a similar value $2.262\pm0.122$
published by BES in Ref.~\cite{Ablikim:2006aj}
and does not contradict to the result of the BES 
measurement~\cite{:2009jsa}:
$R=2.14\pm0.01\pm0.07$ at $W=3.65$~GeV.

\begin{figure}[t!]
\includegraphics[width=0.47\textwidth]{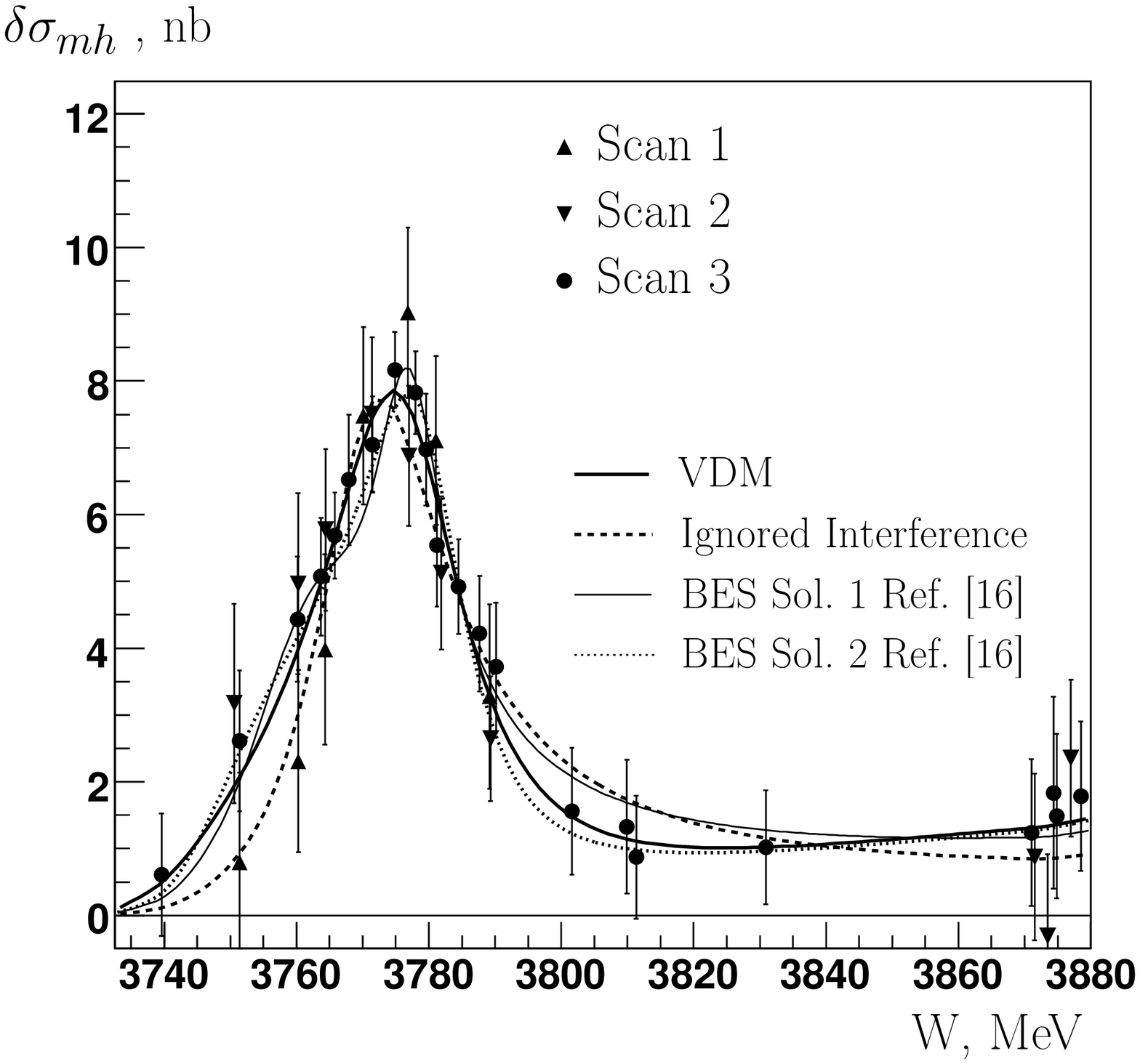}
\caption {{\normalsize \label{PsiPrimeScans_figures}
    Excess of the multihadron cross section 
    in the \DP region. 
    The curves show relevant parts of the fits. The error
    bars correspond to the uncertainty of the measured multihadron
    cross section. All data
    are corrected for the detection efficiency which is different in
    the three scans. See the detailed explanation in the text.
}}
\vspace*{5mm}
\includegraphics[width=0.47\textwidth]{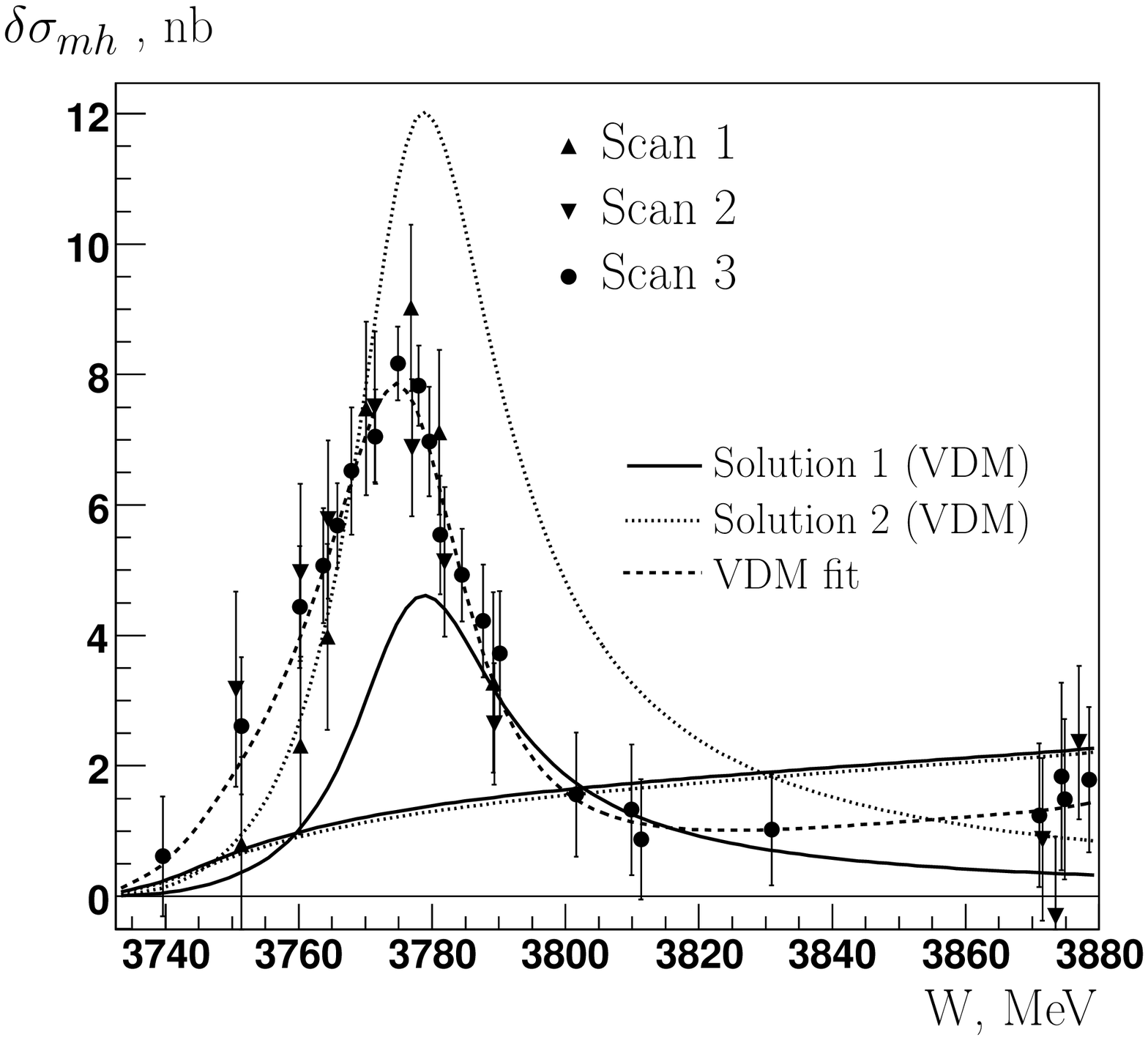}\caption {{\normalsize \label{PsiPrimeScans_solutions}
    Excess of the multihadron cross section in the \DP region.
    Solid and short-dashed curves correspond to two VDM solutions.
    Resonant and  non-resonant parts are presented separately.
}
}
\end{figure}

The excess of the multihadron cross section in the \DP
region is shown in Fig.~\ref{PsiPrimeScans_figures}.
To calculate the excess, the terms 1--4 of Eq.~\eqref{eq:SigMHobs}
obtained by the vector dominance fit were subtracted from the measured cross
section at each point, the residuals were corrected for the 
detection efficiency
calculated by weighting the fit terms 5--8.
These terms of the fits
are presented with the curves. The ignored-interference fit and the fits with
the anomalous line shapes from Ref.~\cite{Ablikim:2008zzcAnShp}
are presented for comparison.

\subsection{On ambiguity of resonance parameters}
It is known that for two interfering resonances 
the ambiguity can appear in the resonance amplitudes
and the interference phase.
A detailed study of that issue can be found in Ref.~\cite{Bukin:2007kx}.
In the case of two resonances with constant widths complete degeneration
occurs: one obtains the identical cross sections for two combinations
of the amplitudes and phase at the same values of the mass and width.

For the energy-dependent widths there is no complete degeneration, however,
the likelihood function has
local maxima on the amplitude-phase plane at slightly
different mass and width values.
A similar situation occurs when a resonance interferes with a varying
continuum.

In our case the typical difference in equivalent $\chi^2$ values
of the two local minima is very small,
$-2\Delta\ln{(L)}\!\simeq\!0.02$, thus a certain solution can not be
chosen. The variation of  mass and width  for possible solutions
is small and neglected below.

\section{Results of analysis}
\subsection{\DP parameters assuming vector dominance}

In Table~\ref{tab:psi3770VDM} we compare the \DP parameters obtained 
under the assumption of \PP dominance in
the nonresonant form factor for two possible solutions  with
those extracted from the ignored-interference fit
and the current world average values. The small corrections
to residual background given below in Table~\ref{tab:FitCor}
of Sec.~\ref{sec:bgcor} are not applied to results of the fit.
The continuum \DD cross section $\sigma^{NR}_{\DD}$ is given without
the radiative correction factor of about 0.75. The values
of the mass and the electron width for the ignored-interference fit
are in good agreement with the world average ones, while the value of
the total width deviates from the average one by 1.5 standard deviation. That
is probably due to the statistical fluctuation that occurred at the
three points of the first scan (see Fig.~\ref{PsiPrimeScans_figures}).

Taking into  account the resonance--continuum interference
in \DD production improves the chi-square of the fits from 91.1/ 73
to 74.8/71. The phase of the \DP amplitude relative to the nonresonant
form factor is about 171 and 240 degrees for the first and second
solution, respectively. The nonresonant form factor has a negative
real part and a small imaginary one. At the \DP peak
\PP contributes approximately 70\% to the total value of the 
nonresonant form factor.
If the resonance-continuum interference
is ignored, the total width 
is not substantially affected,
however, the mass shift of about -6.0~MeV appears as well as 
dramatical  change of the value and  error of the electron width.
The nonresonant \DD cross section in this case
is underestimated as was discussed in the introduction.

A large splitting of the $\Gamma_{ee}$ values is expected
in the near-threshold
region. Let us illustrate that with an example of
the area method of the $\Gamma_{ee}$ determination
discussed soon after the $J/\psi$ discovery~\cite{Jackson:1975vf}.
The electron width is proportional to the area under the resonance curve
\begin{equation}
  \Gamma_{ee} = k\,\frac{M^2}{6\pi^2} \int\!\!\sigma_{res}(W) dW
\end{equation}
(the coefficient $k$ is equal to unity for the energy-independent
total width),
therefore
the following expression can be obtained in absence of radiative corrections
for the case when the continuum
cross section is small compared to the resonant one:
\begin{equation}
\begin{split}
  \Gamma^{\,\text{i.i.}}_{ee} \approx & \:\Gamma_{ee} \left(
    1\!+\!\frac{\alpha}{3} \sqrt{\frac{R_{\text{C}}(M)}{B_{ee}}} \sin{\phi}
  \right)
  +\frac{2\alpha\!\sqrt{B_{ee}}}{3\pi}\,M\cos{\phi} \,\,\times
  \\
    & k\!\!\int\! 
     \frac{(W\!-\!M)\sqrt{\phantom{\!^1\!}\,\Gamma(M)\Gamma(W)\,}
                 \sqrt{\phantom{\!^1\!}\,R_{\text{C}}(W)\,}}
     {(W\!-\!M)^2+\Gamma(W)^2/4}\,\frac{dW}{W}.
\label{eq:ill}
\end{split}
\end{equation}
Here $\alpha$ is the fine structure constant,
$R_{\text{C}}$ is the continuum contribution to $R$,
$B_{ee}$ -- the \ee branching fraction and $\phi$ is the interference
phase. The continuum cross section
$\propto\!(\!\!\sqrt{R_{\text{C\,}}})^2$ is neglected.

The left part of \eqref{eq:ill}  corresponds to the area under 
the measured curve
(\Gee is obtained ignoring the interference), the
right part has  three terms corresponding to the area under 
the resonance curve itself (the true \Gee), the curve due to the imaginary
part of the resonance amplitude (it is also proportional to \Gee) and the
area of the interference wave due to the real part of the amplitude. 
\renewcommand{\arraystretch}{1.4}
\begin{table*}[t!]
\caption {{\normalsize \label{tab:psi3770VDM} \DP fit results 
for the vector dominance
compared to the ignored-interference case.}}
\begin{center}
\begin{tabular}[l]{|c|c|c|c|c|c|c|c|c|} \hline
Solution  & $M$, MeV& $\Gamma$, MeV & $\GeeDP$, ~eV & $\phi$, degrees & $\Gamma^{\PP}_{\DD}$, MeV & $F_0$  & $\sigma^{NR}_{\DD}$, nb & $P(\chi^2)$,\%  \\*[0.2ex] \hline
1 &   $3779.3^{+1.8}_{-1.7}$ &   $25.3^{+4.4}_{-3.9}$ & $ 160^{+78}_{-58} $ & $170.7\pm 16.7$ &$12.9^{+18.5}_{-11.8}$ & $-4.8^{+3.0}_{-3.6}$ & $1.83\pm 0.96$  & $35.7$ \\\hline
2 &   $3779.3^{+1.8}_{-1.6}$ &   $25.3^{+4.6}_{-4.0}$ & $ 420^{+72}_{-80}$  & $239.6\pm 8.6$  &$11.5^{+16.5}_{-10.5}$ & $-4.9^{+3.3}_{-3.7}$& $1.71\pm 0.86$   & $35.7$ \\\hline\hline
i.i. &  $3773.3 \pm 0.5$ & $23.3^{+2.5}_{-2.2}$ & $249^{+25}_{-22}$ & - & - & - & $0.07^{+0.09}_{-0.07}$ & \phantom{$1$}$7.5$ \\\hline
PDG~\cite{PDG} &  $3772.92 \pm 0.35$ & $27.3\pm1.0$ & $265\pm18$ & - & - & - & - & - \\\hline
\end{tabular}
\end{center}
\end{table*}

\renewcommand{\arraystretch}{1.4}
\setlength{\tabcolsep}{3pt}
\begin{table*}[t!]
\caption {{\normalsize \label{tab:psi3770} \DP fits results for alternative assumptions
on the nonresonant form factor $f_D$.}}
\begin{center}
\begin{tabular}[l]{|l|c|c|c|c|c|c|c|c|c|} \hline
Model & \multicolumn{3}{|c|}{Mass, total width and  $P(\chi^2)$ } & \multicolumn{3}{|c|}{ Solution 1 (smaller $\phi$)} &\multicolumn{3}{|c|}{ Solution 2 (larger $\phi$)}  \\\hline
  Equation &$M$, MeV& $\Gamma$, MeV &  $P(\chi^2)$,\%& $\phi$, degrees&$\GeeDP$~, eV& $\sigma^{NR}_{\DD}$, nb&$\phi$, degrees & $\GeeDP$~, eV& $\sigma^{NR}_{\DD}$, nb  \\*[0.2ex] \hline
\eqref{eq:FnrQ} n=1
 &$3779.1^{+2.0}_{-1.6}$ &$24.4^{+5.0}_{-3.6}$& $32.7$ & $167.6\pm 16.0$& $146^{+66}_{-48}$& $1.82  \pm 0.76$ & $243.1\pm\phantom{1}9.5$  & $417^{+75}_{-65}$ & $1.76  \pm 0.73$ \\\hline%


\eqref{eq:FnrQ} n=0.5
 &$3779.0^{+1.7}_{-1.6}$ &$25.5^{+3.0}_{-3.5}$ &  $33.1$& $172.2\pm 17.3$& $172^{+241}_{-66}$& $1.59 \pm   0.86$ & $241.0\pm15.6$  & $418^{+76}_{-65}$ & $1.55 \pm   0.66$ \\\hline%
\eqref{eq:FnrW}
&$3779.0^{+2.1}_{-1.9}$ &$24.4^{+5.1}_{-3.7}$  & $32.7$& $167.5\pm 21.3$&$145^{+83}_{-49}$& $2.09 \pm 0.87$ &$243.1\pm \phantom{1}9.5$  &$418^{+76}_{-74}$ &$2.02 \pm 0.86 $   \\\hline%
\eqref{eq:FnrQW}  n=1
 &$3779.0^{+2.0}_{-1.7}$ &$24.4^{+5.1}_{-3.7}$ & $32.7$ &$167.4\pm 20.4$&$145^{+68}_{-49}$&  $2.14 \pm 0.88$ & $243.0\pm\phantom{1}9.6$   & $422^{+75}_{-74}$ &$ 2.07 \pm 0.86 $    \\\hline%
\eqref{eq:FnrQW}  n=0.5
 &$3779.0^{+1.7}_{-1.6}$ &$25.2^{+4.2}_{-2.8}$& $33.1$ & $172.2\pm 21.6$&$171^{+68}_{-65}$&  $1.81 \pm 0.88$  &$241.3\pm 11.9$  & $419^{+75}_{-68}$ &$ 1.76 \pm 0.85 $    \\\hline%

\eqref{eq:FnrBetaExp} n=0
 &$3779.6 \pm 2.0$ &$25.3\pm 6.6$ &$31.9$ &$200.4\pm 14.7$ & $137 \pm 87$&  $2.20 \pm  0.93$   &$230.3\pm 33.0$  & $461\pm 73$ &$2.47  \pm 1.37$ \\\hline%
\eqref{eq:FnrBetaExp} n=1
 &$3779.6 \pm 1.9$ &$25.3 \pm 6.3$& $31.8$& $176.1\pm 16.6$ & $154 \pm 113$& $2.14   \pm 0.91$ &$239.4\pm 14.7$  & $433\pm 74 $ & $ 1.96   \pm 0.96$  \\\hline%

\eqref{eq:FnrBetaExp} n=3
 &$3779.1 \pm 1.7$ &$25.2 \pm 4.4$&$32.9$&$126.0\pm 15.8$ & $139 \pm 88$& $1.89 \pm 0.90 $ &$ 282.0\pm 16.9$  & $501\pm 89 $ & $ 2.54 \pm 0.91$  \\\hline%
\end{tabular}
\end{center}
\end{table*}

Far enough from the threshold, $R_{\text{C}}$ and $\Gamma(W)$ are almost
constant and the integral is suppressed proportionally
to $\Gamma/M$.
However, for a varying $R_{\text{C}}$ and an asymmetric $\Gamma(W)$
near the threshold, it grows up to
$\,0.02\!\div\!0.15\sqrt{R_{\text{C}}(M)\,}$ depending
on the assumptions about the energy dependence of $\Gamma$ and $R_{\text{C}}$.
The closeness to the threshold increases the influence of the interference
effects by an order of magnitude.
The coefficient preceding $\cos{\phi}$ in Eq.\eqref{eq:ill}
is about 18~keV in the \DP case, the fits give $R_{\text{C}}(M) \simeq 0.3$
with a 40$\div$50\% statistical uncertainty.
Together these circumstances make
the area method inapplicable to \DP.
A fit of the cross section is obviously not so sensitive
to taking interference into account, nevertheless
a splitting of about 260~eV in Table~\ref{tab:psi3770VDM}
does not seem surprising.

The resonant and continuum cross sections 
for the two VDM solutions
are presented in Fig.~\ref{PsiPrimeScans_solutions}. The choice of the true
solution is essential for determination of the non-\DD branching 
fraction of \DP.
At the c.m. energy of 3773 MeV the resonance cross section 
of $3.8^{+1.9}_{-1.4}$~nb
for the first solution and $9.9^{+1.7}_{-1.9}$~nb for the second one should
be compared with the non-\DD cross section, which is $1.08\pm0.40\pm0.15$~nb
according to BES~\cite{Ablikim:2008zzb} and $-0.01\pm0.08^{+0.41}_{-0.30}$
according to CLEO~\cite{Besson:2005hm}. The branching fraction of about 28\% for
the first solution seems unreasonable, however, that can not be considered as
a strong argument in favor of the second solution until improvement in
the non-\DD cross section accuracy.

\subsection{Model dependence of results}
\label{sec:Errors}
To evaluate the model dependence of the \DP parameters
and to check the validity of the vector dominance approach,
the fits were performed with the alternative
assumptions about the nonresonant form factor $f_D(W)$ 
described in Sec.~\ref{sec:Fnr}.
The results of the fits are presented in Table~\ref{tab:psi3770}.
A few other assumptions were also tried.

The amplitude-phase ambiguity was found  in all cases considered. 
For each fit
we assigned the number 1 to the solution with a smaller phase value, 
while the alternative solution got the number 2.
The electron width for the first solution was always smaller than
that of the second one
and the values for two clusters did not overlap.

The results obtained assuming $q^2$ dependence
of the nonresonant form factor as in Eq.\eqref{eq:FnrQ} almost coincide
with those for $W\!-\!m_D$ and mixed dependence in Eq.~\eqref{eq:FnrW}
and \eqref{eq:FnrQW} because of the relatively narrow 
energy range of the experiment.

The mass parameter $a_m$ of the parameterizations
of Eq.~\eqref{eq:FnrBetaExp} $n=0,1,3$ lies between the \PP mass 
and the \DD threshold confirming the \PP dominance.
 Accepting that 
the \DP parameters
corresponding to the vector dominance model are the most reliable, 
we derive the following estimates
for the model dependence: $\delta M = {}^{+0.3}_{-0.3}$~MeV, 
$\delta \Gamma = {}^{+0.2}_{-0.9}$~MeV for both solutions
and $\delta \Gamma_{ee} = {}^{+13}_{-25}$ $\left({}^{+90}_{-10}\right)$~eV,
$\,\delta \sigma^{NR}_{\DD} = {}^{+0.4}_{-0.2}$ $\left({}^{+0.8}_{-0.2}\right)$~nb 
for solutions 1 (2), respectively.
The maximum deviation of parameters from the
VDM results was taken. The definition of the phase $\phi$ with Eq.\eqref{eq:FF}
allows its model-to-model variation, however,
the difference with VDM exceeds the
statistical uncertainty only in the cases \eqref{eq:FnrBetaExp} $n=0,3$
due to a relatively large imaginary part of the nonresonant form factor
fitted in these cases.

We also fitted our data with the anomalous line shapes
considered in the \TCM{Ref.~\cite{Ablikim:2008zzcAnShp} by BES}
where a sum of two noninterfering Breit-Wigner cross sections
and a sum of two destructively interfering amplitudes
were referred to as Solution 1 and Solution 2, respectively.
 The parameters 
of the amplitudes were fixed according to Ref.~\cite{Ablikim:2008zzcAnShp},
the two free parameters were introduced to correct the general normalization
and the shift of the energy scale. The \DP scale correction
averaged for two shapes is $1.042 \pm 0.052$ 
at the energy shift of
$0.92\pm0.51$~MeV
which demonstrates rather good consistency of KEDR and BES data in general.
The chi-square probabilities $P(\chi^2)$ 
are 25.4 and 30.3\% for the solutions
1 and 2, respectively, compared to 35.7\% for the vector dominance fit.
Both shapes provide a better description of the data than
the single Breit-Wigner amplitude not interfering with the nonresonant one
(``i.i''. case in Table~\ref{tab:psi3770VDM})
due to increase of the resonant yield below 3765 MeV. In addition, the 
destructive interference in Solution 2 reduces the resonant yield above
3790 MeV but that does not improve significantly
the general fit quality
because of the growth of the peculiarity in the 3765$\div$3780 MeV energy
region absent in our case. Accounting for the resonance--continuum
interference with a Breit-Wigner resonance amplitude provides 
the best fit of our data although with our statistics
we can not exclude the shape anomaly reported
reported in Ref~\cite{Ablikim:2008zzcAnShp}.
It is worth noting that interference of the \DP structure
with the continuum \DD amplitude should be considered for any
shape assumed.

\subsection{Correction for residual background}
\label{sec:bgcor}

The residual machine background 
is about 2\% of the observed $uds$ cross section
for the scan of 2006 and five time less for the scans of 2004
(Sec.~6.3 of \PrevLet). The estimated numbers of background events 
are $445\pm 97$ and $24\pm7$, respectively, whereas
the total number of multihadron events selected above the \DD threshold
is 33678.

To evaluate the impact of the residual background
 on the resulting fit parameters, the background admixture was
changed in a controllable
way. To do so, we prepared a few samples of background events
passing some loose selection criteria but rejected by the multihadron
ones. At each data point $i$ the number of multihadron events $N^{mh}_i$
was replaced with $N^{mh}_i + f \cdot N^{bg}_i$, where
$N^{bg}_i$ is the number of events in the background sample
The fits with the modified number of events show that
the variations of all fit parameters are proportional to $f$
in the case $ \vert f \vert\, \cdot N^{bg}_i\ll N^{mh}_i$. Selecting
the negative $f$ values at which the total number of subtracted events
matches the  expected background admixture and taking into account 
a small detection efficiency change,
we obtain the corrections for the fit parameters
presented in Table~\ref{tab:FitCor}. The systematic uncertainties
quoted include those of the background admixture estimate and
the variation of corrections obtained using
different background samples.

\begin{table}[h!]
\caption {{\normalsize \label{tab:FitCor} Correction to fit results 
compensating the bias due to the background admixture.}}
\begin{center}
\begin{tabular}[l]{|l|c|c|} \hline
Correction & Solution 1 & Solution 2 \\ \hline
\hspace*{0.5em}$\delta M$, MeV & $-0.06 \pm 0.06$ & $-0.06 \pm 0.06$ \\ \hline
\hspace*{0.5em}$\delta\Gamma$, MeV & $-0.4 \pm 0.3$ & $-0.4 \pm 0.3$ \\ \hline
\hspace*{0.5em}$\delta\Gamma_{ee}$, \% &$-3.9\pm 2.9$ & $-1.5\pm 1.1$ \\ \hline
\hspace*{0.5em}$\delta\sigma^{NR}_{\DD}$, \% &$+1.5\pm 0.5$ & $+1.5\pm 0.5$\\*[0.2ex] \hline
\hspace*{0.5em}$\delta \overline{R}_{uds}^{~2004}$, \% &$-0.5\pm 0.3$ & $-0.5\pm 0.3$ \\ \hline
\hspace*{0.5em}$\delta \overline{R}_{uds}^{~2006}$, \% &$-2.5\pm 1.0$ & $-2.5\pm 1.0$ \\ \hline
\end{tabular}
\end{center}
\end{table}

\subsection{Systematic uncertainties}\label{sec:systerr}

The main sources of systematic uncertainty in \DP parameters 
are listed in Table~\ref{Tab::DPErr}.

\begin{table}[h]
\caption{{\normalsize \label{Tab::DPErr} Systematic uncertainties 
on the \DP mass, total width and electron partial width.
For the latter the uncertainties of two
solutions are presented where different. The 
uncertainty on the nonresonant \DD cross section is also presented.}}
\begin{center}
\begin{tabular}{lcccc} \hline
\setlength{\tabcolsep}{1pt}
{Source}  & \hspace*{-0.5em}$M$[MeV]  &\hspace*{-0.2em}$\Gamma$[MeV]
       &\hspace*{-0.7em}$\Gamma_{ee}$[$\%$] &\hspace*{-0.7em}$\sigma^{NR}_{\DD}$[\%]\\*[0.2ex]  \hline \hline 
 \multicolumn{4}{c}{Theoretical uncertainties and external data precision} \\ \hline      
 $\mathcal{B}_{n\DD}$   & $^{+0.0}_{-0.5}$  & $^{+0.0}_{-0.2}$
        & \hspace*{-0.5em}$^{+8.8}_{-0}\big/\phantom{}^{+0}_{-2.3}$ & ${}^{+0}_{-12.}$ \\*[0.2ex] \hline
 $R_{0}$ value in $\Gamma(W)$    & 0.3  & 0.3 &  2. & 1.5 \\  \hline
 $\Gamma_{\DnDn}/\Gamma_{\DpDm}$ & 0.1  & 0.1 & 0.4 & 0.8\\ \hline
 $D,\overline{D}$ masses & \phantom{$1$}0.06 & \phantom{$1$}0.04  &  \phantom{$~$}0.3  & 0.5\\  \hline
 $\DD\pi$ cross section  & \phantom{$1$}0.15 & \phantom{$1$}0.05  & 1. & 2. \\  \hline
\multicolumn{4}{c}{Detector and accelerator related uncertainties} \\ \hline
  Det. efficiency variation \hspace*{-1.5em} 
                   & 0.03  & 0.04
                                      & 2.4  &5.\\  \hline
 Hadronic event selection \hspace*{-1.em}
                         & 0.3  & 0.3 & 3. & 5. \\  \hline
 Residual background
                         &0.06  & 0.3
                                        & 2.9 &3. \\\hline
 Luminosity measurement  & 0.1  & 0.1 & 2.  & 2. \\  \hline
 Beam energy             &  \phantom{$1$}0.03 & --  & -- & -- \\         
\hline
\hline
\emph{Sum in quadrature}   &   $^{+0.48}_{-0.69}$  &  $^{+0.54}_{-0.58}$  &
    \hspace*{-0.5em}$^{+10.5}_{-5.7}\big/\phantom{}^{+5.7}_{-6.1}$
    & $^{+8.}_{-14.}$\\*[0.2ex] \hline
\end{tabular}
\end{center}
\end{table}
When the resonance-continuum interference is
taken into account, the multihadron cross section becomes rather
sensitive to the non-\DD fraction of \DP decays.
It was varied from zero to 0.16 
as was mentioned
in Section~\ref{sec:fitting}. The variations of the \DP mass and total
width were 0.3 and 0.1~MeV, respectively. The shift of the electron width
was +8.8\% for the first solution and -2.3\% for the second one.

The uncertainty on the $R_{0}$ value used to specify the energy-dependent
width \eqref{eq:GammavVSw}, \eqref{eq:BWdampF}
of about $25\%$ (\TCM{Refs.~\cite{Buchmuller:1981,Godfrey:1985}})
leads to these of $0.3$~MeV both in the mass and  total width.
When the interference is ignored, the sensitivity to $R_0$
variations reduces by a factor of 3.

The \TCM{uncertainties} due to \TCM{that} of the branching fraction ratio 
for \DnDn and \DpDm 
are about 0.1~MeV for the mass and total width.
Approximately the same uncertainties are obtained because of the $D$ meson 
masses. The estimates were obtained
by variation of the values within their errors quoted by PDG.

To estimate \TCM{uncertainties} due to the
\TCM{inaccuracy of the} $\DD\pi$ cross section \TCM{treatment}
at the edge of the energy range of the experiment, we used two methods: 
shrinking of the fit range and assumption of the linear dependence 
on the $D$-meson c.m. velocity  instead of the cubical one 
in Eq.~\eqref{eq:CMSbeta}. The latter corresponds
to variation of the effective interaction radius $R_0$ for
$\DD\pi$ states from zero to infinity.
The variations of the mass, total width and electron width do not
exceed $0.15$~MeV, $0.05$~MeV and $1\%$, respectively. 

The systematic uncertainties due to the energy dependence
of the detection efficiencies shown in Table~\ref{tab:def} can be neglected
in all cases except $\epsilon_{uds}$. The latter together with the
energy dependence of the radiative correction factor and
possible $R_{uds}$ variation determine
the power in the expression \eqref{eq:1s} used to parameterize
the light quark contribution to the multihadron cross section. The
radiative correction factor
$1+\delta^{RC}_{uds}=1.125\pm0.022$ was calculated according to
Ref.~\cite{KF} using
the vacuum polarization data compilation by the CMD-2 group reviewed
in Ref.~\cite{Actis:2010gg}. The error quoted includes
the uncertainty of the detection efficiency dependence on the mass
of the hadronic system produced via ISR and that of the vacuum
polarization data.
We explicitly considered the $J/\psi$
tail in the cross section~\eqref{eq:SigMHobs}, thus the correction
factor is $14\div9$\% less than that used in 
\TCM{Ref.~\cite{Ablikim:2006aj}}
and its variation in the experiment
energy range does not reach 0.1\%. The precise $R$ measurements
at $W=3.07$ and $3.65$~MeV~\cite{:2009jsa} do not indicate essential
$R_{uds}$ variation, thus we concluded that
the $uds$ efficiency variation dominates in the
uncertainty of the power $1\!-\!\delta$. Performing the fits with
different values of $\delta$ we evaluated the uncertainty
of the \DP parameters as 0.03~MeV, 0.04~MeV and 2.4\%
for the mass, total width and electron width, respectively.
Compared to that, the energy dependence of $\epsilon_{\DD}\,$
gives only a 0.5\% bias of the electron width and a few keV
shifts of the mass and total width.

The sensitivity of the mass and width to the criteria of the multihadron
event selection
was checked by changing cuts on the energy deposited in the 
calorimeter and conditions on 
the number of tracks. The results were stable within 0.3~MeV.
The detection efficiency uncertainty due to inaccuracy
of the $D$-meson decay ratios~\cite{PDG} used for the simulation contributes
$2\%$ to the electron width \TCM{uncertainty}. The dependence on the choice of
the selection criteria increases it up to 3\%. The sensitivity
to the event selection criteria is partially due to the
influence of the residual background. We ignore that and
treat the background correction as an independent \TCM{uncertainty} source
which makes the \TCM{uncertainty} estimates more conservative.

 Uncertainties due to the luminosity measurement instability
are less than $0.1$ MeV for the mass and width. The accuracy
of the absolute luminosity measurements discussed
in \PrevLet\,  contributes less
than $2\%$ to the electron width \TCM{uncertainty}.
The uncertainty on \DP mass due to the beam energy determination 
does not exceed $30$~keV.

\section{Summary}
The parameters of the \DP meson  have been measured using the data 
collected with the KEDR detector at the VEPP-4M $e^+e^-$ collider.
Interference of resonant and nonresonant production essential in the
near-threshold region has been taken into account.

Our final results on the mass and width of \DP are: \\ [-3pt]
\begin{equation*}
 \begin{split}
  M & =  3779.2\,\,^{+1.8}_{-1.7}\,\, ^{+0.5}_{-0.7} \,\,^{+0.3}_{-0.3} \,\, \text{MeV}, \\
 \Gamma & = \:\:\:\:\,24.9\,\,^{+4.6}_{-4.0}\,\, ^{+0.5}_{-0.6}\,\,^{+0.2}_{-0.9} \,\,    \text{MeV,} \\      
 \end{split}
\end{equation*}
 
The corrections applied to the fit results are listed in Table~\ref{tab:FitCor}.
The third error arises from the model dependence. It was estimated
comparing the results obtained under the assumption of
vector dominance in the $D$-meson form factor (quoted values)
and under a few alternative assumptions which
do not imply vector dominance.
The quoted model errors do not include possible
deviations of the resonance shape from the Breit-Wigner one
with usual assumptions about the
total width energy dependence, which are predicted, e.g., in the
coupled-channel model~\cite{Eichten:1980}.

The result on the \DP mass agrees with that by
BaBar also taking into account interference (Ref.~\cite{BABARDATA})
and is significantly higher than
all results obtained ignoring this effect.
The mass values obtained studying $B$-meson decays by 
BaBar~\cite{Aubert:2007rva}
and Belle~\cite{:2007aa} are lower but do not contradict to our
measurement.

We got two possible solutions for the \DP electron partial 
width and the radiatively corrected nonresonant \DD cross section
 at the mass of \DP :
\begin{equation*}
  \begin{split}
& (1)  \:\:~\GeeDP = 154\,^{+79}_{-58}\,^{+17}_{-9}\,^{+13}_{-25}\: \text{eV},
     \:\:~\sigma^{NR}_{\DD} = 1.4\pm0.7\,\,^{+0.1}_{-0.2}\,\,^{+0.3}_{-0.2}
       \:\text{nb}, \\
& (2)  \:\:~\GeeDP= 414\,^{+72}_{-80}\,^{+24}_{-26}\,^{+90}_{-10}\: \text{eV},
     \:\:~\sigma^{NR}_{\DD} = 1.3\pm0.7\,\,^{+0.1}_{-0.2}\,\,^{+0.6}_{-0.2}
       \:\text{nb}. \\
\end{split}
\end{equation*}
The phase shifts of the \DP amplitude relative to the negative nonresonant
amplitude are $171\pm17$ and $240\pm9$ degrees for solutions (1) and (2),
respectively.

Most of potential models support the first
solution and can barely tolerate the second
one. The increase of the \DP mass according to the 
BaBar and KEDR
measurements implies the decrease of the $2S$-$1D$ mixing
used in potential models to rise the electron width value above 100~eV
(Refs.~\cite{Badalian:2009,Radford:2007,Gonzalez:2003,Richard:1980}
and the reviews~\cite{Brambilla:2011,Brambilla:2004wf}). The correct choice 
of the true solution is extremely important for a determination 
of the non-\DD fraction of
\DP decays.

Because of the large uncertainty the solution (1) does not contradict formally
to the previously published results, which 
do not take the interference effect into account, the solution
(2) is only two standard deviations higher than the current world average.
However, the qualitative consideration
and numerical estimates confirm that the impact of the resonance--continuum
interference on the resulting electron width value is
large, therefore the resonance parameters obtained 
taking into account interference  
can not be directly compared with the corresponding values obtained 
ignoring this effect.

\section*{Acknowledgments}
We greatly appreciate permanent support of the staff of the 
experimental, accelerator and electronics
laboratories while pre\-paring and performing
this experiment. Stimulating discussions with V.~P.~Druzhinin
are acknowledged.

This work was partially supported  by the Ministry of Education and
Science of the Russian Federation and grants Sci.School 6943.2010.2,
RFBR 10-02-00695, 11-02-00112, 11-02-00558.

\end{document}